\documentclass[11pt,twoside]{article}
\usepackage{asp2010}

\resetcounters

\begin{document}

\title{Imaging the circumstellar dust distribution around AGB stars with the NOT/PolCor instrument}
\author{Sofia Ramstedt$^1$, Matthias Maercker$^{1,2}$, G\"oran Olofsson$^3$, Hans Olofsson$^{3,4}$ and Fredrik L. Sch\"oier$^4$
\affil{$^1$Argelander Institut f\"ur Astronomie, Auf dem H\"ugel 71, DE-53121 Bonn, Germany}
\affil{$^2$European Southern Observatory, Karl Schwarzschild Str. 2, Garching bei M\"unchen, Germany}
\affil{$^3$Department of Astronomy, SE-10691 Stockholm, Sweden}
\affil{$^4$Onsala Space Observatory, Dept. of Radio and Space Science, Chalmers University of
Technology, SE-43992 Onsala, Sweden}}

\begin{abstract}
Stellar light from an AGB star is scattered by the circumstellar dust and polarized in the direction perpendicular to the source. Therefore, images of circumstellar envelopes around AGB stars in polarized light traces the dust distribution and can be used to search for asymmetries, and to achieve a better understanding of the mechanisms at play when AGB stars are transformed into planetary nebulae. The PolCor instrument is a combined imager, polarimeter, and coronograph providing images with an angular resolution down to 0\farcs2. We have used it to map the dust distribution around three AGB stars: W~Aql, and the detached shell sources DR~Ser, and U~Cam. W~Aql is a binary and we find indications of a bi-polar dust distribution around the star. The observations of the latter two sources clearly reveal the detached shells, likely the result of a brief, strongly enhanced mass-loss rate during the late evolution of these stars. Mapping the detached shells gives us clues to the symmetry of the mass loss and important evolutionary processes.
\end{abstract}

\section{The PolCor instrument and the observations}
The PolCor instrument is a combined imager, coronograph and polarimeter built for the Nordic Optical Telescope (NOT). PolCor provides sharp images (resolution down to 0\farcs2), has a well-defined PSF (resulting in a higher image contrast) and a high-quality polarimeter.

\begin{figure}[t]
\begin{center}
\includegraphics[width=5.1cm]{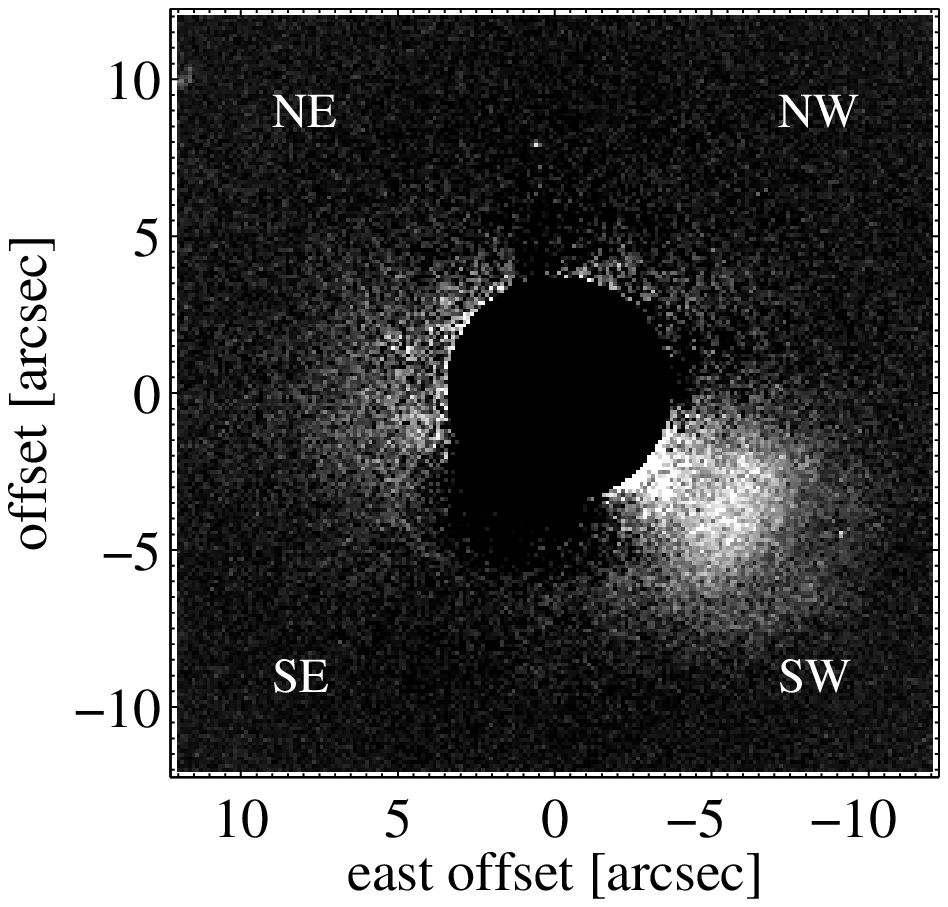}
\includegraphics[width=5.1cm]{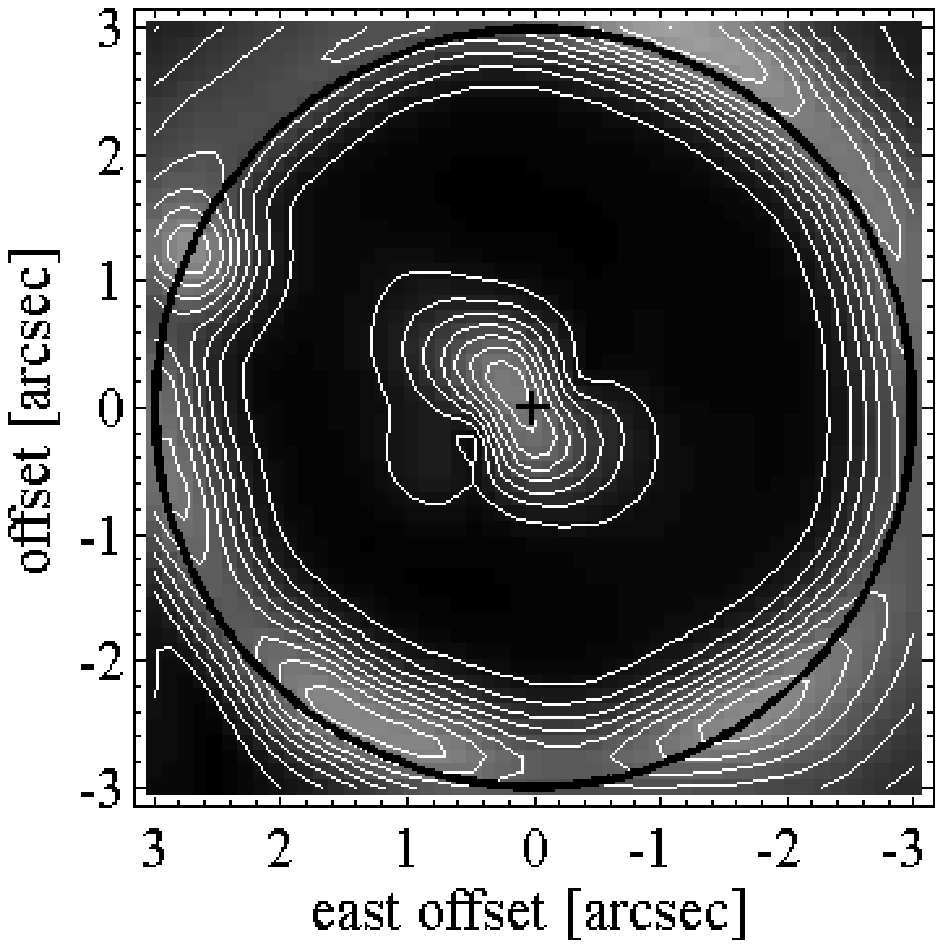}\\
\caption{{\em Left:} The large-scale image of the polarized light around W~Aql. To enhance asymmetries a Gaussian has been subtracted from the image. {\em Right:} Contour map of the polarized light in the inner 3'' as seen through the absorbing coronographic mask. The cross marks the position of the AGB star and the ring the edge of the mask. }
\label{waql}
\end{center}
\end{figure}

We observed three sources using PolCor/NOT: the binary S-type AGB star W~Aql, and the two detached shell carbon stars U~Cam and DR~Ser. Observations of the polarized light around the stars trace the circumstellar dust distribution and thus enables the search for structures and asymmetries in the circumstellar envelope. Though this study the geometry of the mass loss can be investigated which will give important clues not only to the processes at play, but also to the transition from the AGB phase to the planetary nebula phase.

\section{Results}
The circumstellar dust distribution around the binary star W Aql is mapped and found to be asymmetric, both on large (10'') and on smaller (1'') scales (Fig.~1). The large-scale images show what appears to be a dust-density enhancement on the south-west side of the star and the inner region appears bi-polar through the coronographic mask. These structures are in agreement with what could be expected from binary interaction. 

The detached shells are clearly seen in the images (Fig.~2). Measurements of their width and radius agree with previous investigations. 

For further details on the results see Ramstedt et al. (2010, {\em submitted to A\&A}).

\begin{figure}[h]
\begin{center}
\includegraphics[width=5.1cm]{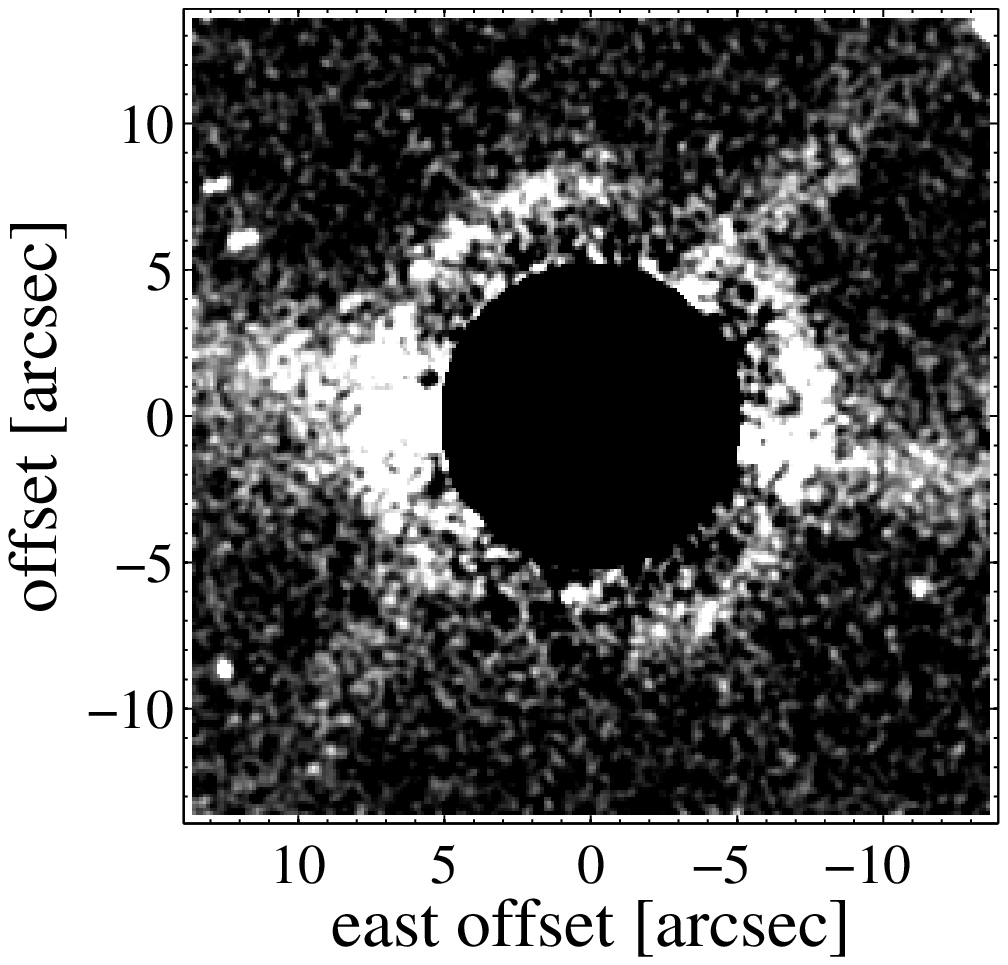}
\includegraphics[width=5.1cm]{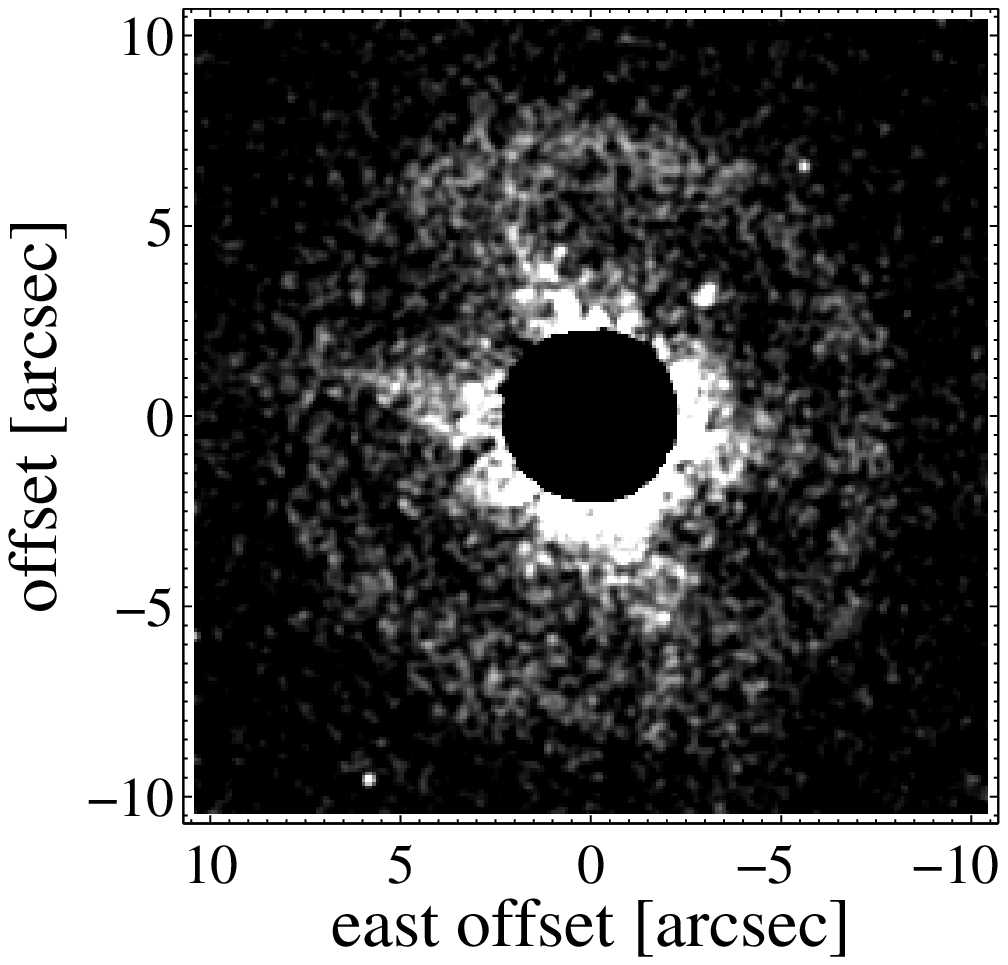}
\caption{{\em Left:} The detached shell around U~Cam. {\em Right:} The detached shell around DR~Ser.}
\label{ful}
\end{center}
\end{figure}

\acknowledgements SR acknowledges support by the Deutsche Forschungsgemeinschaft (DFG) through the Emmy Noether Research grant VL 61/3-1.


\end{document}